\documentclass[letterpaper, 10pt, conference, dvipsnames]{ieeeconf}  

\IEEEoverridecommandlockouts                              

\usepackage{lineno}   
\usepackage[american]{babel}
\usepackage{mathtools} 
\usepackage{booktabs} 
\usepackage{tikz} 
\usepackage{tkz-graph}
\usetikzlibrary{calc, shadows, plotmarks, shapes,arrows,
  decorations.pathmorphing, arrows.meta,bending,matrix,positioning}
\tikzstyle{naveqs} = [text width=40mm, fill=ForestGreen, minimum height=20mm, rounded corners]
\tikzstyle{ann} = [above, text width=5em]

\usepackage{algorithm}
\usepackage{algorithmic}
\usepackage{graphicx}
\usepackage{subcaption}
\usepackage{caption}
\usepackage{amsmath,amssymb,amsfonts}
\usepackage{soul}
\usepackage{mathtools}

\usepackage{float}
\usepackage{cite}
\usepackage{textcomp}

\usepackage{xcolor}

\makeatletter
\let\NAT@parse\undefined
\makeatother
\usepackage{hyperref}  

\title{\LARGE \bf
Decentralized Traffic Flow Optimization Through Intrinsic Motivation
}

\author{Himaja Papala$^{1}$, Daniel Polani$^{2}$, and Stas Tiomkin$^3$\\
$^1$Charles W. Davidson College Of Engineering, San Jose State University, CA, USA\\
$^2$School of Physics, Engineering and Computer Science, University of Hertfordshire, UK\\
$^3$Computer Science Department, Whitacre Collage of Engineering, Texas Tech University, Texas, USA\\
Emails: himaja.papala@sjsu.edu, d.polani@herts.ac.uk, stas.tiomkin@ttu.edu\\
{\bf Published in:} 2024 IEEE 27th International Conference on Intelligent Transportation Systems (ITSC)\\
\url{https://ieeexplore.ieee.org/document/10920268}
\thanks{$^{3}$ S. Tiomkin (Corresponding Author) was with Charles W. Davidson CoE, San Jose State University, when most of this work was done.}%
}

\begin{document}
\maketitle
\thispagestyle{empty}
\pagestyle{empty}

\begin{abstract}
Traffic congestion has long been an ubiquitous problem that is exacerbating
with the rapid growth of megacities. 
In this proof-of-concept work we study \emph{intrinsic motivation}, implemented via the empowerment principle, to control autonomous car behavior to improve traffic flow.
In standard models of traffic dynamics, self-organized traffic jams emerge spontaneously from the individual behavior of cars,
affecting traffic over long distances. Our novel car behavior strategy improves traffic flow while still being decentralized and using only locally available information without explicit coordination. Decentralization is essential for various reasons, not least to be able to absorb robustly  substantial levels of uncertainty. 

Our scenario is based on the well-established traffic dynamics model, the \emph{Nagel-Schreckenberg cellular automaton}. In a fraction of the cars in this model, we substitute the default behavior by empowerment, our intrinsic motivation-based method. This proposed model significantly improves overall traffic flow, mitigates congestion, and reduces the average traffic jam time. 
\end{abstract}


\section{Introduction}\label{sec:Intro}

Traffic congestion poses a significant challenge, with its impact increasingly felt on the global economy. Among the various solutions for traffic management, centralized approaches \cite{yamashita2004car, chen2007reliable, zambrano2019centralized} are typically restricted to areas with high-quality infrastructure, enabling near real-time connectivity to central servers and neighboring vehicles. This is a
limiting assumption, in terms of location and of
the number of cars. 

We therefore assert that any viable scalable solution must be decentralized, eliminating the dependence on a central hub or protocol between vehicles, and must be capable of withstanding significant levels of uncertainty and noise. This opens the door to artificial intelligence approaches that hold promise in making traffic more efficient.

A priori, the  drivers' individual goal would be to both safely and quickly arrive at their destination. However, there is no obvious externally-defined objective for artificial agents which would simultaneously aim to achieve the collective good of maintaining high-throughput traffic flow and mitigate the congestion level. Our proposal is to use a suitable intrinsic motivation measure to control the individual cars, which lead to the emergence of global collective behaviour with less traffic jams.

\emph{Intrinsic motivation} has been increasingly established as a class of artificial agent behavior strategies that do not rely on a specific reward \cite{charlesworth2019intrinsically, guckelsberger2016intrinsically, hrl2016, choi2021variational, gregor2016variational}, but, inspired by biological agents, are driven by objectives  internal to
the agents or via their interaction with the environment. Empowerment constitutes a paradigm dual to that
of training artificial agents by externally-provided
objectives in form of rewards or costs which
require expert domain knowledge.

\emph{Empowerment} measures the potential influence of the agent through its
actions on its future; in practice, this may be only partially observable, so observing this influence is limited by the agent's sensors, reflecting the agent's limited environmental knowledge. Empowerment has been demonstrated in a broad range of domains \cite{du2020ave, zhao2020efficient, Empowerment1, Empowerment12, volpi2020goal,salge2014changing,salge2014empowerment, polani2009information, klyubin2005empowerment,tiomkin2024prx} including assistance, locomotion, stabilization, tool use, and others. For our best knowledge it has not been explored for decentralized traffic flow optimization, which is one of the main contributions of this work.    

The present work is a proof-of-concept study of autonomous cars driven by empowerment to improve traffic flow in the collective. We employ the well-established, but highly useful abstracted cellular automaton model for traffic, the \emph{Nagel-Schreckenberg} model \cite{nagel1992cellular,ketaren2020modeling,mori2012aircraft,lin2012optimizing,chen2018effects}, which we will refer to as NaSch. It is widely used as a standard model of traffic dynamics, and it captures effects of traffic, such as the spontaneous emergence of traffic jams affecting traffic over long distances. So, the default behavior is modeled by the 
NaSch model, our intrinsic motivation is realized via \emph{empowerment}.

Our method operates in fully decentralized manner and the autonomous cars use only locally available information, without an explicit coordination or communication protocol even between neighbouring drivers. Such decentralization is essential for any practical scalable solution that ensures drivers' privacy as well as the viability of the method when communication between cars is impaired, e.g., under  network communication failures or lag and without relying on  compatible protocols. 

We address this by replacing, for randomly selected cars, the NaSch rules of the car behaviour with empowerment-based rules, which operates on the individual level.
In the autonomous car, empowerment implements the drive to locally maintain or increase the car's degrees of freedom as much as possible (in terms of local "freedom of operation", e.g.\ moving into states with more options to accelerate or brake) \cite{salge2014empowerment}. It will turn out that this, at the same time, manages to affect the general good favourably. 

We want to emphasize that our goal is not to surpass specific state-of-the-art algorithms or to compare against a diverse set of road optimization models. We acknowledge the importance of such comparisons for full application in traffic scenarios, which we defer to future work. Our work represents the first attempt, to our knowledge, to utilize intrinsic motivation for decentralized traffic optimization. We believe that demonstrating its efficacy in such a diverse scenario is a significant and opens new research directions in hybrid models of intelligent traffic control. 

Our proposed model demonstrates substantial improvements in  traffic flow in most cases with respect to the unmodified original system, even when only a fraction of the cars has been modified to be driven by empowerment.

\section{Prior Work}

We now review existing methodologies for traffic regulation and/or congestion alleviation, specifically in light of the unique characteristics of the present work, namely: i) level of decentralization and inter-vehicle communication, ii) inherent uncertainty concerning traffic state owing to the  restricted (local) information accessible to each car, and iii) the expert domain knowledge typically necessary to formulate effective reward functions for artificial agents. 

The recent work by \cite{vinitsky2023optimizing} on traffic flow
control emphasizes the importance of decentralized autonomous control
and has, to some extent, similar  objectives to the current work. The model defined three types of state spaces, {\it minimal, radar and aggregate} differing in their information granularity.
However, \cite{vinitsky2023optimizing} admit a certain
level of centralization and communication such as requiring 
global information about "the number of vehicles in the bottleneck",
even in the {\it minimal state}, (cf., Section 3.3 in \cite{vinitsky2023optimizing}). Two
other state representations, {\it radar and aggregate},
require GPS information and the average car speed at various road
segments, respectively.
Also, the global reward function must be shared between all the agents
in the network, which requires centralization and network access. 

In \cite{kesting2008decentralized}, decentralized control of traffic is achieved by explicit communication between cars via network. Specifically, when an Inter-Vehicle Communication (IVC) equipped car enters a traffic jam, a corresponding message containing jam related information (position, time) is constructed and broadcasted. However, if there is significant distance between IVC equipped cars or interference due to variations in driving speeds, the communication would become unreliable. 
In \cite{maske2019large}, a decentralized traffic control model is
proposed where agents are distributed across the road and observe
specific areas. Agents provide control commands like desired headway and lane choices to Autonomous Vehicles (AVs) within their respective regions. The decentralized control is achieved by allowing agents' observational spaces to overlap while keeping their controlling spaces non-overlapping, facilitating an understanding of other agents' behavior without direct communication. However, communication between agents and AVs is still necessary in this model. In \cite{kim2021congestion}, a congestion-aware collaborative automatic cruise control is proposed, which requires vehicle-to-everything technology (V2X) for inter-vehicle communication.  In \cite{zhang2023learning}, \cite{cui2021scalable}, Multi
Agent Reinforcement Learning (MARL) is used to achieve
distributed control with local information and no inter-agent
communication. While these approaches align with our goal,
they face the challenging task of having to design reward
functions to train the agents.

In contrast to the existing works, we propose a fully decentralized approach for traffic regulation without any type of inter-vehicle communication, with a substantial uncertainty about the overall traffic state, and without manually designed reward function. 

\section{Preliminaries}
\subsection{Traffic Model}

\noindent The standard Nagel–Schreckenberg
(NaSch) model \cite{nagel1992cellular} which we use here  is a stochastic particle
hopping model used to simulate traffic flow. It  matches real-life
traffic patterns to a remarkable extent and  predicts traffic flow phenomena with remarkable accuracy. The model captures complex phenomena observed in real traffic, including traffic jams,
and it is an important tool for the analysis of traffic flow \cite{nagel1994life, nagel1995emergent}.

In the NaSch model one considers a road divided into $L$ cells and typically with periodic boundary conditions. Each cell can either be occupied by one of $N$ cars or left empty. The car density on this road is thus given by $\rho=\frac{N}{L}$. 

\noindent{\bf Traffic Dynamics.} In the  NaSch model, each car has not only a location, but a positive velocity along the direction of movement. These are discrete integers ranging from $v_{\text{min}}=0$, indicating an idle car, to a predefined maximum speed $v_{\text{max}}>0$ that a car can move at. The velocity of a car $i\in\{1\dots N\}$ in the cell $x_i$ and at time $t$ is denoted by $v(x_i, t)$.

The traffic dynamics in the NaSch model is defined by four update
rules, $R_1,\dots, R_4$, which are applied sequentially to all the cars  at each
time step, $t\in[1,\dots, T]$, with $T$ denoting the total number of
updates (the simulation time):

\begin{enumerate}
    \item[$R_1$]{\bf(Accelerate):} Increase the velocity $v(x_i, t)$
      of the car in cell $x_i$ by $1$ if it is less than the maximum velocity $v_{\text{max}}$.
    \begin{equation} \label{rule1}
        v = \min(v+1, v_{\text{max}})
        \end{equation}
    \item[$R_2$]{\bf(Brake to avoid collisions):} If the velocity
      $v(x_i, t)$ of a car at cell $x_i$ is greater than the distance,
      $\Delta\doteq x_{k}-x_{i}$, to the next car at cell $x_k$ (i.e.\
      if acting upon it would cause a collision with the front car), then decrease $v(x_i, t)$ to $\Delta-1$.
     \begin{equation} \label{rule2}
         v = 
        \begin{dcases}
            \Delta-1, & \text {if } v \geq \Delta\\
            v,   &\text{otherwise}
        \end{dcases}
    \end{equation} 
    \item[$R_3$]{\bf (Random Brake):} Reduce the velocity $v(x_i, t)$ of a car by $1$ with a probability $0<p_{brake}<1$.
    \begin{equation} \label{rule3}
         v = 
        \begin{dcases}
            v-1, & \text {if } v > 0 \text { }\& \text { } \mathtt{rand()}< p_{brake}\\
            v,   &\text{otherwise}
        \end{dcases}
    \end{equation}
    \item[$R_4$]{(\bf Move):} Advance each car by $v(x_i, t)$ cells according to its velocity.
\end{enumerate}

These simple rules capture plausible driver behavior, such as the
desire to drive as fast as possible, while acknowledging that
acceleration is limited {\it(Rule 1)}, the inclination to avoid
collisions {\it(Rule 2)} and the influence of stochasticity due to
various factors such as inconsistent human driving behavior, weather
or road conditions, as well as other external influences {\it (Rule 3)}.
Finally, {\it (Rule 4)} describes the forward movement of the vehicles.

Traffic simulated by these rules shows several essential features of
real traffic, including regions of free traffic flow, traffic
congestion, and phase transitions taking place between these two modes
of movements. The validity of the core model is well-established in
\cite{lubeck1998density,roters1999critical}. Notably, these
rules not only directly govern the behavior of individual cars, but
rather give raise to intricate collective phenomena such as long-range
traffic waves.

\begin{figure}[t!]
\centering
\includegraphics[width=0.9\columnwidth]{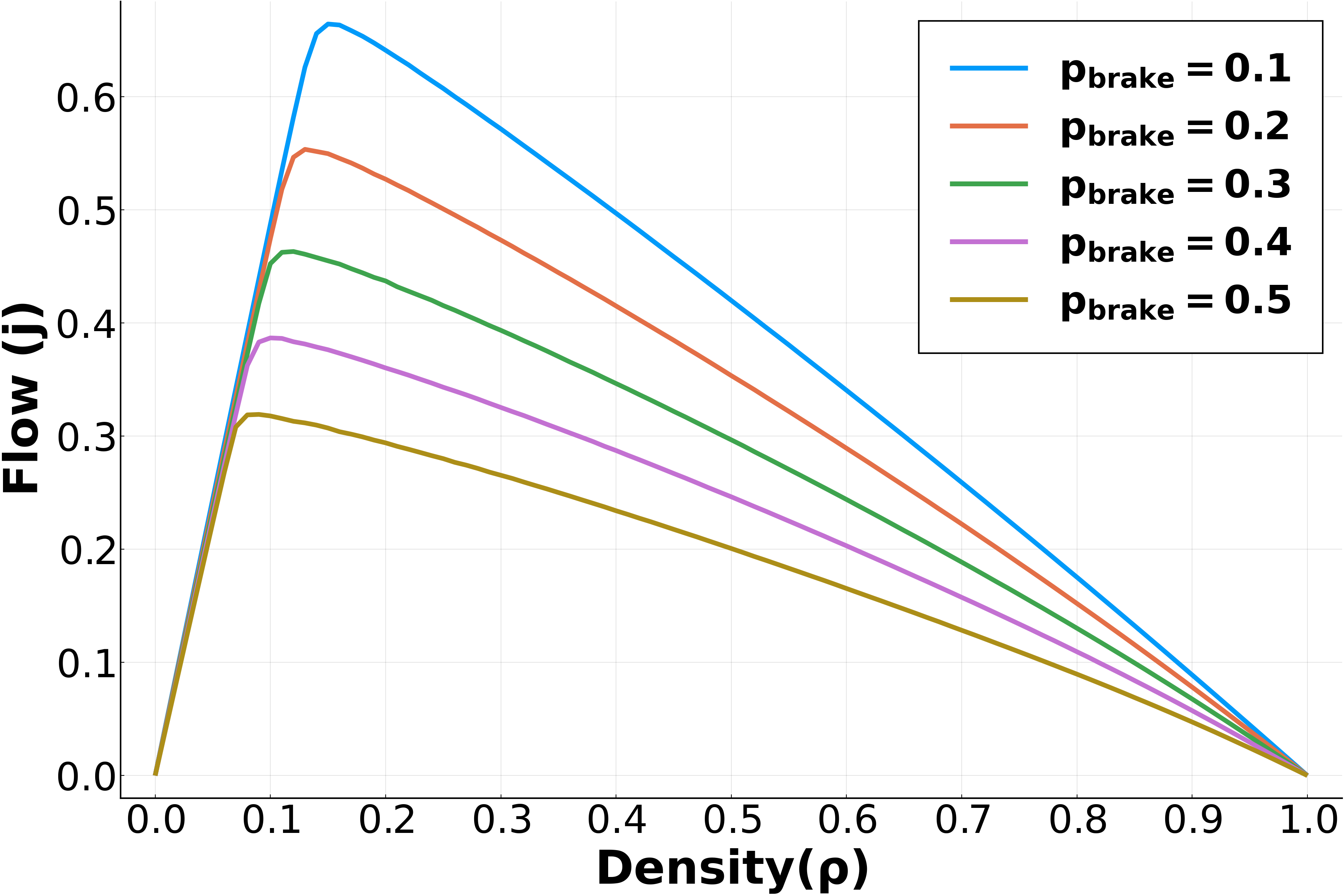} 
\caption{{\bf Fundamental diagram} of expected traffic flow for different values of the braking probability $p_{brake}$.}
\label{fundamental-diagram-different-p-brake}
\end{figure}

\noindent{\bf Fundamental diagram of traffic flow.} The main quantity in the NaSch model is the traffic flow, $j$. It measures how many cars pass through a given spot in a
given time  \cite{nagel1992cellular}. 
The expected traffic flow, ${j}$, can be estimated for a finite number of simulation steps, $T$, and a finite number of cells, $L$ with $N$ number of cars, as follows:
\begin{equation}
    {j} \approx \frac{1}{L}\frac{1}{T}\sum_{i=1}^{N}\sum_{t=1}^{T}v(x_{i},t).\label{eq:traffic flow approx}
\end{equation}

Figure~\ref{fundamental-diagram-different-p-brake} illustrates the
dependence of the traffic flow as a function of car density, also
known as the {\it fundamental diagram of traffic flow}. This diagram
exhibits essential characteristics of traffic flow, namely a steep
increase for low traffic densities, representing {\it free-flowing traffic}.
The flow increases linearly as cars can travel at full speed without
any forced deceleration. Subsequently, we see at a specific density,
(the {\it critical density} $\rho_c$), an abrupt transition to a regime, $\rho > \rho_c$,
where average flow decreases due to
reduced space, increased braking and emergence of self-organized traffic jams. This
diagram is commonly used for the analysis of the traffic properties
 and the peak of the diagram indicates the maximally reachable
traffic flow; the higher the curve lies in the fundamental diagram, the
better the flow achieved\footnote{The simulation is conducted on a road with a length, $L$, of $100,000$ cells, $v_{\text{max}}=5$ and $T=5000$. To ensure accurate data collection, the initial $1000$ time steps are disregarded to allow transient effects to stabilize. Subsequently, data is collected at every $5^{th}$ time step to reduce correlations between consecutive time steps.}.

The objective of the current work is to develop a fully decentralized method that enables
to increase the traffic flow over a broad range of car densities on
the fundamental diagram by adding a small fraction of agents driven by intrinsic motivation to the traffic. These agents neither communicate with each other nor do they coordinate by
explicit protocols. Instead, they follow a variation of the NaSch rules,
enhanced by the \emph{empowerment} measure.

As we will show, this is sufficient to  globally improve traffic, which is characterized by a new \textit{augmented fundamental diagram} wherein the peak is shifted upwards and towards the right. 

\subsection{Intrinsic Motivation by Empowerment }

We propose to cast the problem of mitigation of traffic congestion as
a problem of achieving {\it traffic flow viability and self-maintenance} for the individual car agent: traffic jams decrease this viability and harm
autonomous flow maintenance for that care. Concretely, we suggest empowerment as a
suitable candidate for the intrinsic objective for the mitigation of
traffic congestion. It has been demonstrated by various studies to
provide artificial agents with the desired properties of viability and
drives towards self-maintenance \cite{  zhao2020efficient, Sharma2020Dynamics-Aware, salge2017empowerment, du2020ave}, remarkably often coinciding with behaviors induced by explicit hand-designed reward functions.

Formally, one characterizes empowerment as the {\it channel capacity} between a potential sequence of agent
actions and the (possibly limited) agent's sensor in the future after executing this sequence \cite{klyubin2005all}. Empowerment is defined for a given time horizon of actions. Concretely, for a $n$-step
action sequence $A^n$, define $n$-step empowerment as\footnote{We denote random variables by capital letters, and their particular realisations by small letters.}:
\begin{align}
\mathcal{E}^n(s_t) =&\max_{p(A^n_{t}\mid s_t)}I[S_{t+n};A^n_{t}\mid s_t],\label{eq:emp}
\end{align}
where $I[S_{t+n};A^n_{t}\mid s_t]$ is the mutual information between the distribution of final states, $S_{t+n}$, and that of the action sequence, $A^n_{t}$, conditioned on the specific current state, $s_t$. Stated differently, it is the channel capacity between $A^n_{t}$ and $S_{t+n}$. We emphasize that $A^{n}_{t}$ is a distribution over \emph{potential} actions, not the action actually taken by the agent --- the latter is selected to move into a state with maximum such channel capacity, i.e.\ one maximizing the term from \eqref{eq:emp}.

The essential properties of empowerment are summarized as follows. When the agent's set of
actions only lead to outcomes that are similar or closely aligned,
the agent's capacity to exert influence is constrained, reflected
through low values of empowerment.
Conversely, when the available action spectrum results in a wide range
of potential distinguishable states, the agent's ability to shape its
environment is considered stronger; this corresponds to larger
empowerment values. 
In our traffic scenario, when a car agent is trapped closely behind e.g.\ a slow-moving car, and there are not
many different state outcomes the current agent can impose, empowerment will
be low \cite{salge2014empowerment} and vice versa. 

\begin{figure*}[ht!]
   \centering
   \begin{tabular}{cccc}
   \hfill
    \begin{subfigure}{0.32\linewidth}
        \includegraphics[width=\linewidth]{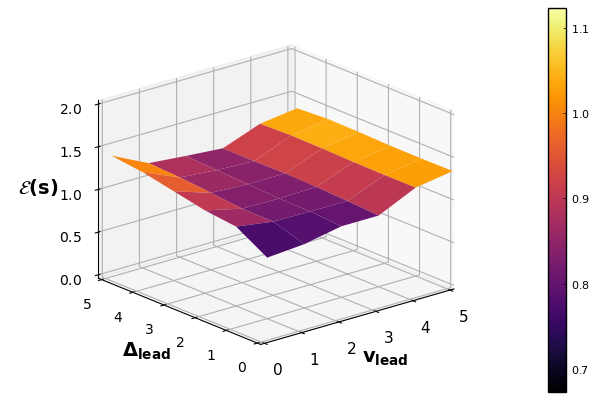}
        \caption{$v_{agent}=0$}\label{Es_va0}
    \end{subfigure}
    \hfill
    \begin{subfigure}{0.32\linewidth}
        \includegraphics[width=\linewidth]{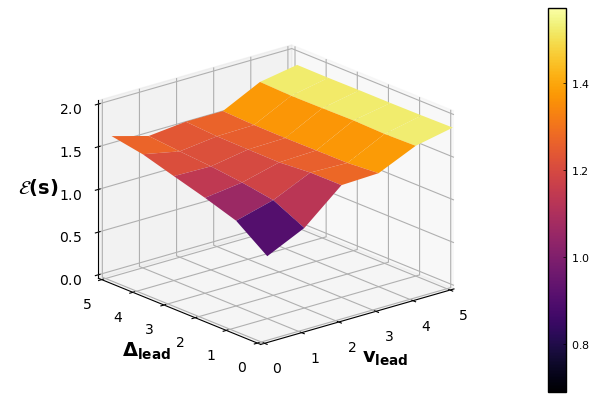}
        \caption{$v_{agent}=2$}\label{Es_va2}
    \end{subfigure}
    \hfill
     \begin{subfigure}{0.32\linewidth}
        \includegraphics[width=\linewidth]{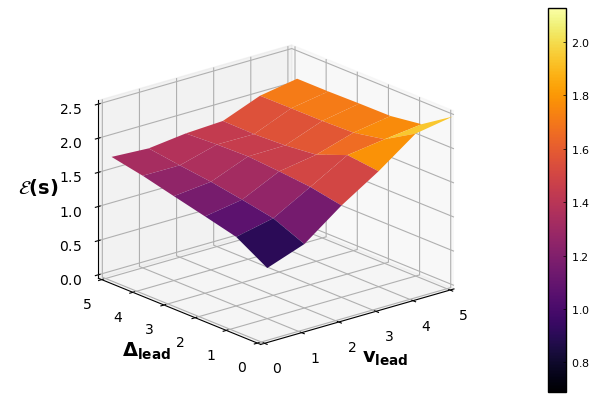}
        \caption{$v_{agent}=5$}\label{Es_va5}
    \end{subfigure}
    \end{tabular}
    \caption{3-step empowerment, $\mathcal{E}^3(s)$, in bits for $p_{brake}=0.2$ and $\rho=0.2$   }
   \label{E(s) vs vn vs dn}
\end{figure*}

\section{Proposed Method}
Here we present our novel approach to employ empowerment
 within the context of the NaSch model and to
investigate its effect on the traffic dynamics. We equip certain cars
with non-default decision-making capabilities throughout the
simulation. These agent cars are not
bound by NaSch rules; instead, they select velocities based on
maximizing their empowerment. We hypothesize that when the cars aim towards states with higher empowerment, this enhances the viability of traffic and decreases traffic jams. 

Starting in a state $s_{t-1}$ (we describe the state in more detail below), each agent computes \eqref{eq:emp problem def} for the possible states $s_t$ it can reach, subject to NaSch dynamics, and then takes the action that moves it into the state with the highest value of, $\mathcal{E}^n(s_t)$: 
\begin{align}
\mathcal{E}^n(s_t)=&\max_{p(a^n_{t}\mid s_t)}I[S_{t+n};A^n_{t}\mid s_t]\label{eq:emp problem def}\\
&\mbox{subject to NaSch dynamics}.\nonumber
\end{align}

Given that a particular car's velocity on the road is primarily influenced
by its leading car's velocity, we focus solely on the local
information with respect to the leading car. The state $s$ of an agent thus encompasses only locally observable information
such as the distance between the agent and its leading car and the
velocity of the leading car $s=( \Delta_{\text{lead}}, v_{\text{lead}})$.  With this local information, agents strive to make decisions, particularly choosing velocity, with the greedy goal of stepping into a state of as high empowerment as possible.  

We consider choosing a velocity as an action $a$, and a sequence of actions $a^n$ representing consecutive actions taken by such an agent over $n$ time steps. Agents have the capacity to increase their velocity by 1, while they can decrease it by any arbitrary amount. Consequently, the one-step actions available to the agent are $a_t \in  \{0,\dots,\min(v_{agent}(t)+1, v_{max})\}$.  Furthermore, we assume that each agent
has a full model of the system rules, including the probability of
braking ($p_{brake}$) and the maximally permissible velocity ($v_{max}$) on the road. Additionally, agents assume their neighboring cars follow traditional NaSch update rules. 
However, the empowered agents still operate under constraints such as the
ability to increase their velocity only by one unit or having to decelerate to avoid
collisions.

The empowerment horizon, the duration for which the
potential future action sequences are probed is a crucial parameter of the
simulation which we set at the outset and which remains constant throughout
the simulation. For each experiment, the overall braking
probability $p_{brake}$  and density $\rho$ is specified. Given this, the transitional probability of velocities for normal cars   $p(v_{t+1}|v_t)$\footnote{ The transitional probability of velocities for normal cars $p(v_{t+1}|v_t)$ have been sampled independently for a given $p_{brake}$ and vehicle density $\rho$, on a lane comprising $10,000$ cells over the course of $10^6$ time steps.} is calculated. To estimate empowerment, agents need to generate the distribution of destination states ($s_{t+n}$) for every available action sequence ($a^n$) --- we call this distribution as state transitional channel $p(s_{t+n}|a^n,s_t)$. With the help of $p(v_{t+1}|v_t)$, agents can construct $p(s_{t+n}|a^n,s_t)$. 

The agents possess knowledge of the transitional probability of the velocity of normal cars $p(v_{t+1}|v_t)$. For each action $a$ in the action sequence, agents sample the $v_{\text{lead}}(t+1)$ (future velocity of their leading car) from $p(v_{t+1}|v_t)$. Subsequently, they can estimate the distance in next time step using $\Delta_{\text{lead}}(t+1) = \Delta_{\text{lead}}(t) + v_{\text{lead}}(t+1) - a$. This process provides the agent with the future sample state $s_{t+1} = (\Delta_{\text{lead}}(t+1), v_{\text{lead}}(t+1))$. Agents iterate through this procedure for each action $a$ in the available action sequence $a^n$, constructing the distribution $p(s_{t+n}|a^n,s_t)$.
Once $p(s_{t+n}|a^n,s_t)$ is constructed, the empowerment of a
particular state is calculated using the Blahut-Arimoto
(BA) algorithm \cite{blahut1972computation} for the channel
$p(s_{t+n}|a^n,s_t)$. 

Figure \ref{E(s) vs vn vs dn} illustrates the interplay between $\Delta_{\text{lead}}$, $v_{\text{lead}}$, and $v_{\text{agent}}$ in influencing the empowerment of a state $\mathcal{E}(s)$. The analysis indicates that larger distances to the lead car $\Delta_{lead}$ generally improve the agent's empowerment. However, the influence of faster velocity of the lead car $v_{lead}$ is stronger, and this effect is more pronounced when the agent's car itself moves faster. This observation aligns logically with the expectation that when agent is trapped behind a slow moving car, there are not many different state outcomes the agent
can impose, because whatever action it takes it has to reduce the velocity to avoid collisions, resulting in low empowerment values. However, if agent is behind the fast moving car, it would have more options to accelerate and can end up in different states resulting in high empowerment values.
 
The main goal of the agent is to find an action that results in more
empowered state. To achieve this, it assesses how empowered it could
be in various possible future situations if it takes a particular
action $a_t$ at a certain time $t$. There are different ways to approach
this, but for the present  purpose, the agent considers the
probability of each outcome for the given one step action $p(s_{t}|s_{t-1},a_{t-1})$ and averages the
empowerment across the outcomes $s_{t}$ for a given action $a_{t-1}$. 
\begin{equation}
  \label{eq:average-action-empowerment}
\langle\mathcal{E}\rangle(s_{t-1}, a_{t-1})\doteq\sum_{s_{t}}p(s_{t}|s_{t-1}, a_{t-1})\mathcal{E}^n(s_{t})
\end{equation}

For simplicity, we omit the current state in the notation, denoting empowerment of the current state, $s_{t-1}$, and the action, $a_{t-1}$, by $\langle\mathcal{E}\rangle$(a). 
Agents always chooses the action $a_{t-1}$ maximizing the expected empowerment in \eqref{eq:average-action-empowerment}. If multiple actions have same expected empowerment, the action is chosen randomly with equal probability.

\begin{algorithm}[ht!]
\caption{Expected Empowerment}
\label{alg:algorithm}
\textbf{Input}: velocity of an agent ($v_{agent}$), initial state ($s_{t-1}$), $p(v_{t+1}|v_t)$, and $v_{max}$ \\
\textbf{Parameter}: Empowerment horizon, $n$\\
\textbf{Output}: $ \langle \mathcal{E}\rangle(a)$
\begin{algorithmic}[1] 
\STATE Initialize 1-step actions $a_{t-1}\in \{0,\dots, \min(v_{agent}+1, v_{\text{max}})\}$\\
\STATE Compute $p(s_{t}|s_{t-1},a_{t-1})$\\
\FOR{$s$ in $s_{t}$}
\STATE Get $n$-step action sequences $a^n$
\STATE Compute channel $p(s_{t+n}|a^n,s)$
\STATE Compute Empowerment, $\mathcal{E}^n(s)$, by Eq. (\ref{eq:emp})
\ENDFOR
\STATE $ \langle\mathcal{E}\rangle(a)\gets \sum_{s_{t}}p(s_{t}|s_{t-1},a_{t-1})\mathcal{E}^n(s_{t}) $\\

\STATE \textbf{return} $\langle\mathcal{E}\rangle(a)$
\end{algorithmic}
\end{algorithm}

\begin{algorithm}[ht!]
\caption{Traffic Flow with Empowerment}
\label{alg:algorithm2}
\textbf{Input}: position of agents ($\widehat{X}$), position of regular cars ($\widehat{X_n}$), road length $(L)$, simulation time $(T)$. \\
\textbf{Parameter}: density, $\rho$, braking probability $p_{brake}$\\
\textbf{Output}: Flow $j$
\begin{algorithmic}[1] 
\STATE Let $t=1$.\\
\WHILE{$t \leq T $}
\FOR{ $x$ in $\widehat{X}$ }
\STATE $v(x,t+1)$= $\underset{a}{\mbox{argmax }} \langle \mathcal{E}\rangle(a)$\\
\STATE $v(x,t+1) \gets \mbox{NaSch Rule 2} $ \\
\ENDFOR

\FOR{ $x_n$ in $\widehat{X_n}$ }
\STATE $v(x_n,t+1) \gets \mbox{NaSch Rules 1-3}$\\
\ENDFOR

\STATE Displace cars by NaSch Rule 4\\
\ENDWHILE
\STATE $j \approx \frac{1}{L}\frac{1}{T}\sum_{i=1}^{N}\sum_{t=1}^{T}v(x_{i},t)$\\

\STATE \textbf{return} $j$
\end{algorithmic}
\end{algorithm}

\noindent{\bf Traffic Flow with Empowerment:}
In the final step, we outline the procedure for calculating traffic
flow in the presence of empowered agents. We begin by randomly
initializing a road of length $L$, populated with a given number of
cars $N$. These cars will have velocities in the range
$v \in \{ 0 : v_{\text{max}}\}$. 
Among these cars, certain cars are considered  empowered agents. Their positions
are recorded in the vector $\widehat{X}$, while 
normal cars' positions are recorded in vector $\widehat{X_n}$. These data
structures are then used for subsequent empowerment
calculations. For each agent in $\widehat{X}$ we will independently
compute the expected empowerment $\langle\mathcal{E}\rangle(a)$ and asynchronously carry out
the action that maximizes $\langle\mathcal{E}\rangle(a)$. After the actions are chosen, agents verify whether their chosen actions result in any collisions. If so, agents adjust their actions to $\Delta-1$ (NaSch rule $R_2$).
All the normal cars in $\widehat{X_n}$ will follow the NaSch update rules ($R_1$, $R_2$,
$R_3$). Subsequently,
all cars, including agents, will follow rule $R_4$ and move forward
based on their velocities. The positions all cars will be
correspondingly updated and stored for the subsequent time step
computations. This process is repeated for $T$ time steps. Upon
completion of the simulation, the traffic flow will be evaluated using
the approximation provided in \eqref{eq:traffic flow approx}.

\captionsetup[subfigure]{justification=centering}
\begin{figure*}[ht!]
   \centering
   \begin{tabular}{cccc}
   \centering
    \begin{subfigure}{0.3\linewidth}
        \includegraphics[width=\linewidth]{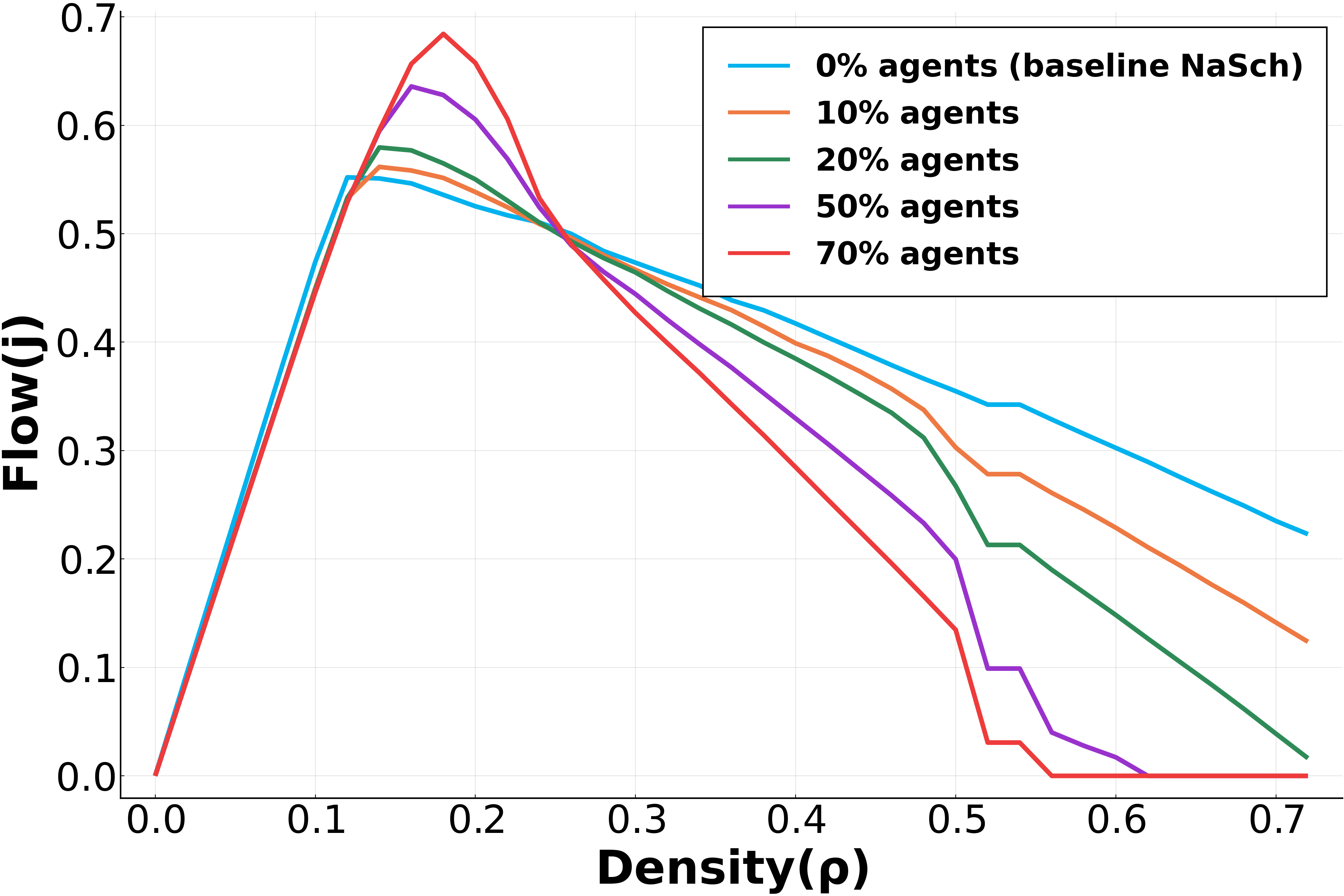}
        \caption{\\$p_{brake}=0.2$\\2-step planning horizon} \label{fd_2step_0.2bp}
    \end{subfigure}
    \centering
    \begin{subfigure}{0.3\linewidth}
        \includegraphics[width=\linewidth]{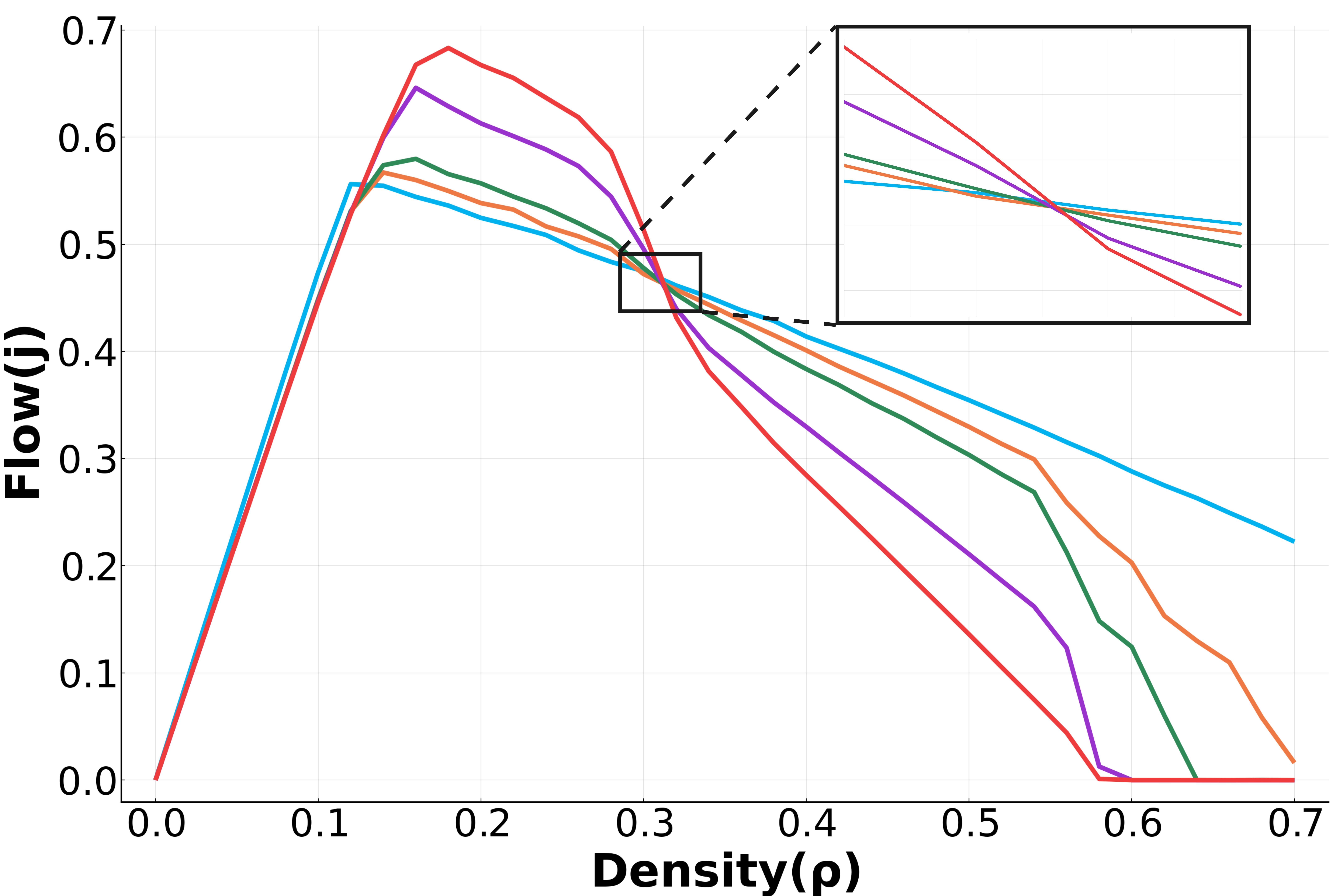}
        \caption{\\$p_{brake}=0.2$, \\3-step planning horizon}\label{fd_3step_0.2bp}
    \end{subfigure}
    \centering
    \begin{subfigure}{0.3\linewidth}
        \includegraphics[width=\linewidth]{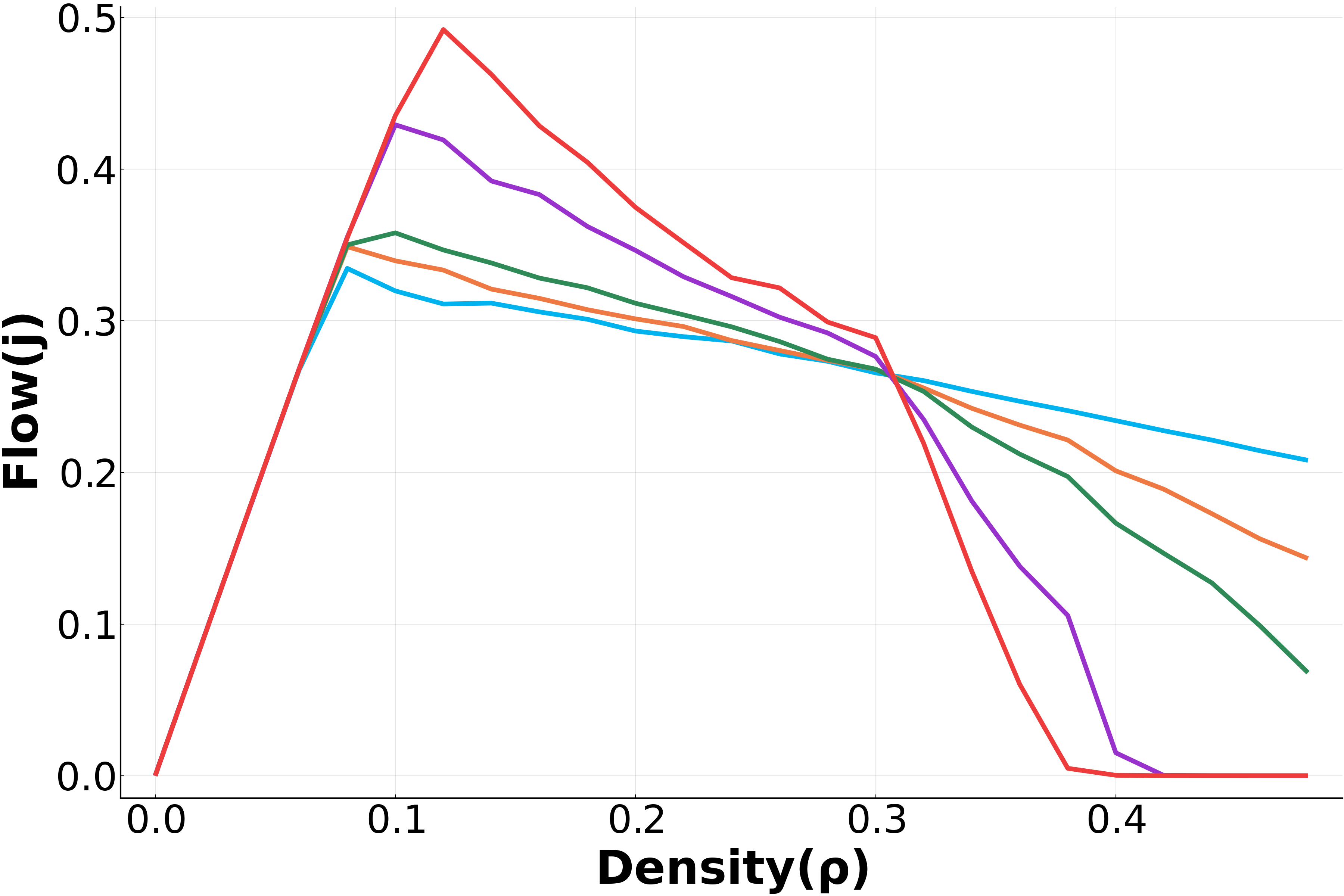}
        \caption{\\$p_{brake}=0.5$, \\3-step planning horizon}\label{fd_3step_0.5bp}
    \end{subfigure}
    \end{tabular}
    \caption{Comparison of traffic flow with varying ratios of empowered agents, $0\%$ to $70\%$, across different vehicle densities on the x axis. It is evident that an increase in the horizon expands the range of densities for which the model remains effective. As expected, with an increase in $p_{brake}$ value, the flow decreases, but the effect of empowerment remains consistent. The analysis is done on a road of $L=1000$, $v_{max}=5$ with $T=5000$. Note, the curves intersect at nearby but not exactly the same point (cf., insert 'b') }
   \label{traffic flow with emp}
\end{figure*}

\begin{figure*}[ht!]
   \centering
   \begin{tabular}{cccc}
   \begin{tabular}{cccc}
   \hfill
    \begin{subfigure}{0.3\linewidth}
        \includegraphics[width=\linewidth]{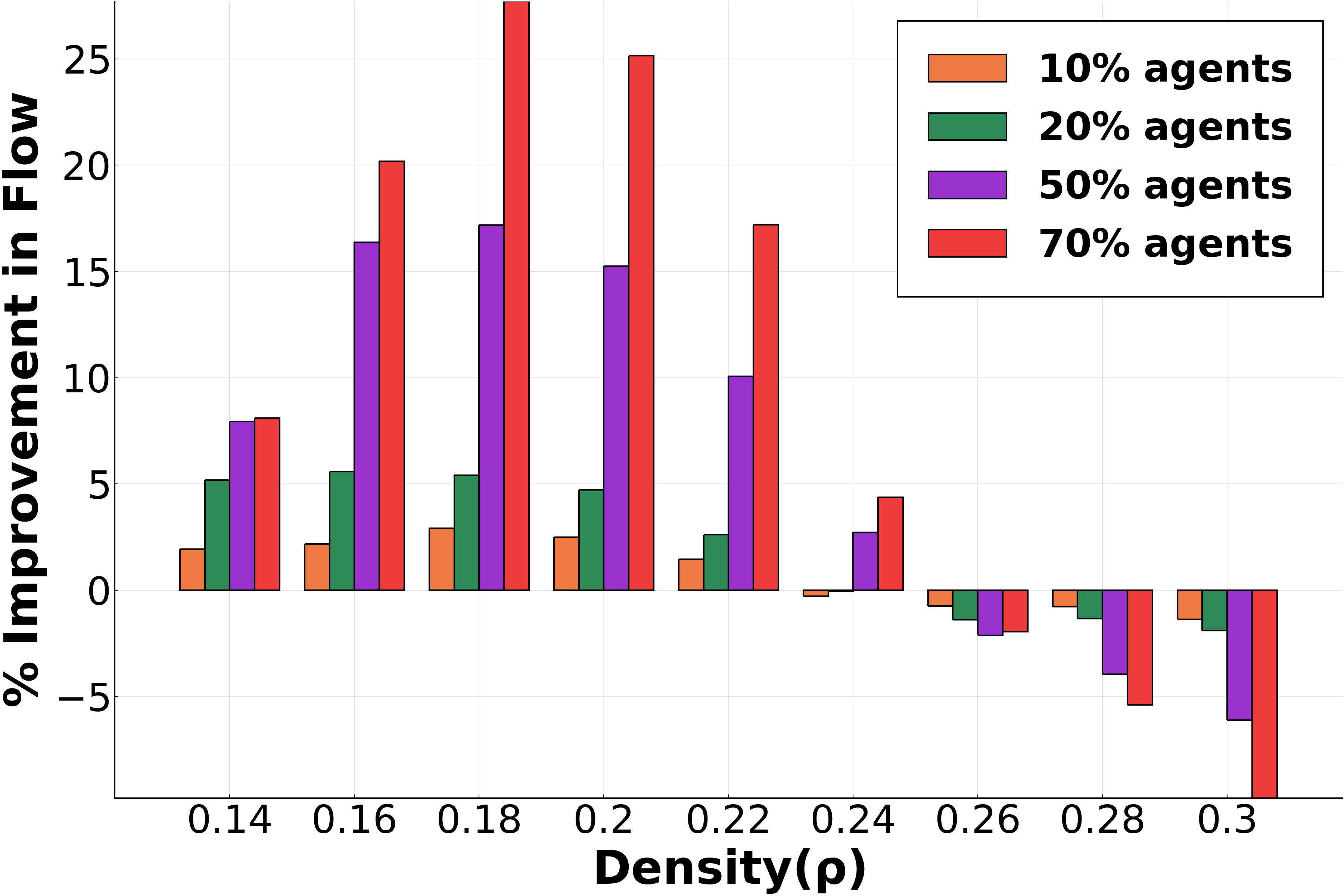}
        \caption{\\$p_{brake}=0.2$, \\2-step planning horizon}\label{flow_imp_2step_0.2bp}
    \end{subfigure}
    \end{tabular}
    \centering
    \begin{tabular}{cccc}
    \hfill
    \begin{subfigure}{0.3\linewidth}
        \includegraphics[width=\linewidth]{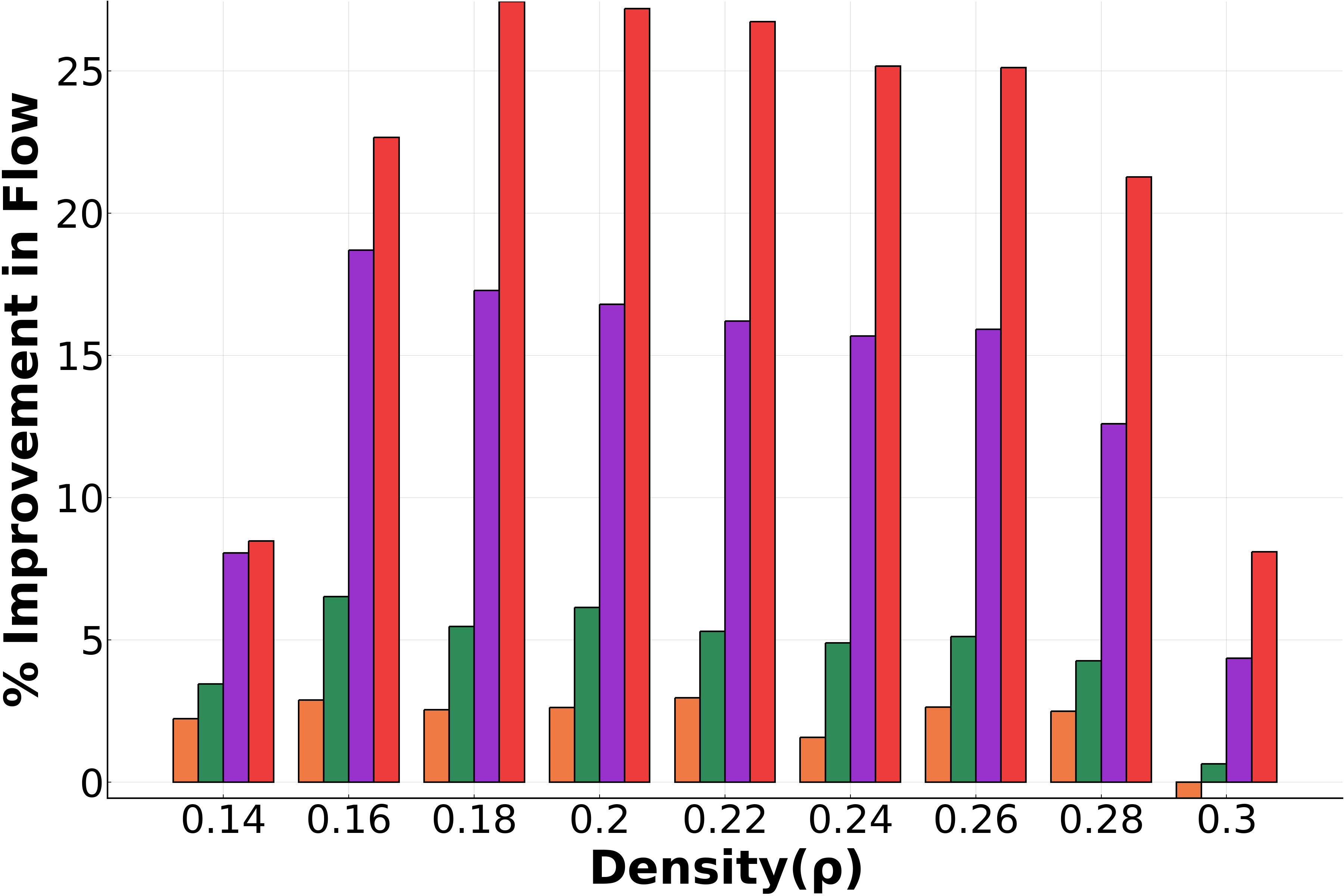}
        \caption{\\$p_{brake}=0.2$,\\ 3-step planning horizon}\label{flow_imp_3step_0.2bp}
    \end{subfigure}
    \end{tabular}
    \centering
    \begin{tabular}{cccc}
    \hfill
    \begin{subfigure}{0.3\linewidth}
        \includegraphics[width=\linewidth]{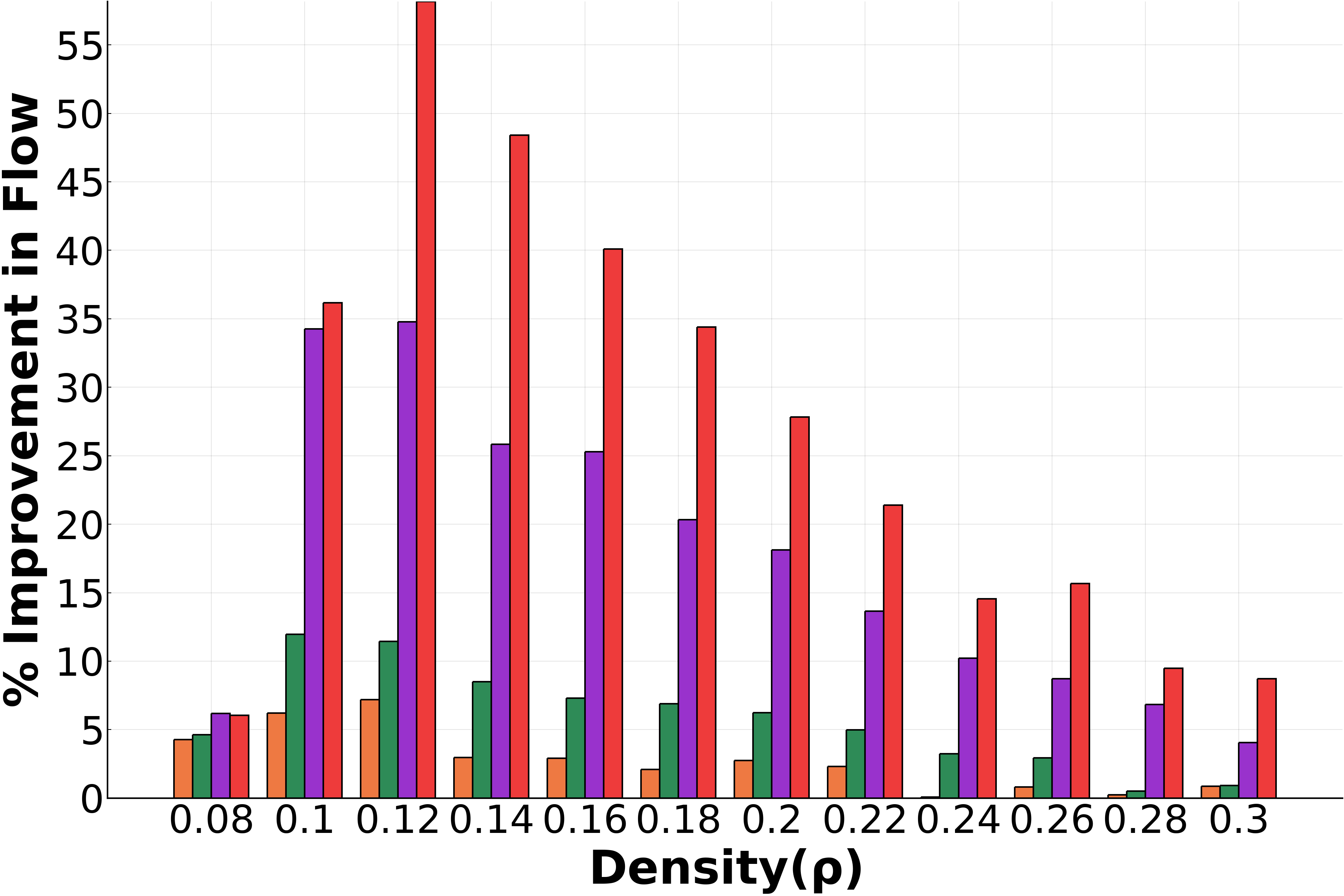}
        \caption{\\$p_{brake}=0.5$,\\ 3-step planning horizon}\label{flow_imp_3step_0.5bp}
    \end{subfigure}
    \end{tabular}
    \end{tabular}
    \caption{ Comparison of the percentage (\%) improvement in traffic flow for densities beyond critical density $(\rho_c)$ and those within the model's effective range, under varying ratios of empowered agents, different $p_{brake}$ values, and different empowerment planning horizons.}
   \label{percentage improvement in flow}
\end{figure*}

\section{Experiments}
We now study the influence of the empowered agents on traffic dynamics and especially its important global signatures, such as traffic flow and average traffic jam time. The full code repository is provided at \cite{codeRepo}. 

In our investigation, we emphasize that agents rely solely on local information for decision-making. We conduct experiments with varying ratios of empowered agents relative to the total number of vehicles on the road. Figure \ref{traffic flow with emp} shows the fundamental diagram illustrating traffic flow across different densities, considering braking probabilities of $p_{\text{brake}}=0.2$ and $0.5$, and two different empowerment horizons.

Remarkably, as depicted in Figure \ref{percentage improvement in flow}, the inclusion of empowered agents yields substantial improvements in traffic flow, reaching approximately $27\%$ for $p_{\text{brake}}=0.2$ and approximately $58\%$ for $p_{\text{brake}}=0.5$, compared to scenarios devoid of empowered agents. These enhancements are observed across a range of densities, extending beyond the critical density $(\rho_c)$.

As the number of agents on the road increases, we observe a corresponding increase in the maximum attainable traffic flow. Additionally, the critical density $(\rho_c)$ shifts rightward, indicating that empowered agents prolong the transition from a free-flowing state to the onset of traffic congestion when compared to the  NaSch model.

\begin{figure*}[h!]
    \centering
    \begin{tabular}{ccc}
        \begin{tabular}{ccc}
            \begin{subfigure}{0.3\linewidth}
                \includegraphics[width=1.05\linewidth]{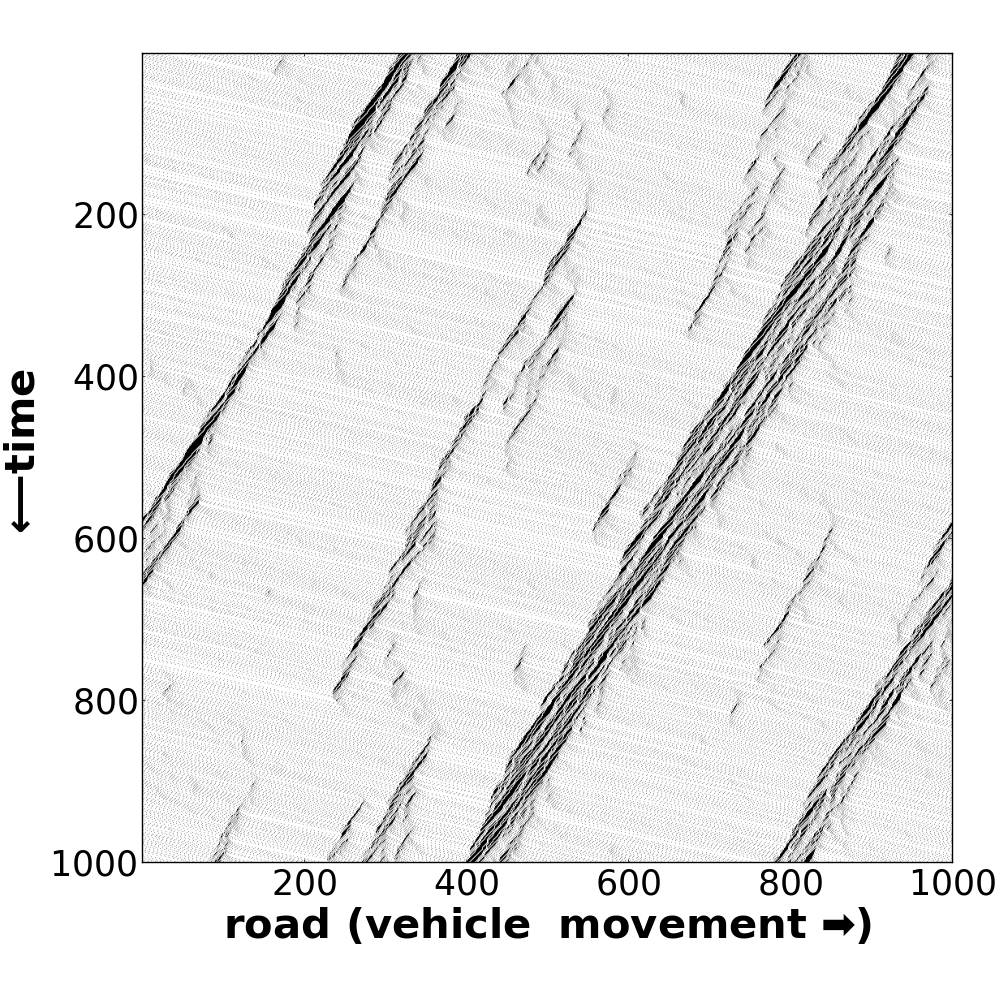}
                \caption{$p_{\text{brake}}=0.2$, $\rho=0.18$,\\NaSch baseline}
                \label{0agents_0.2}
            \end{subfigure}
            &\quad
            \begin{subfigure}{0.3\linewidth}
                \includegraphics[width=\linewidth]{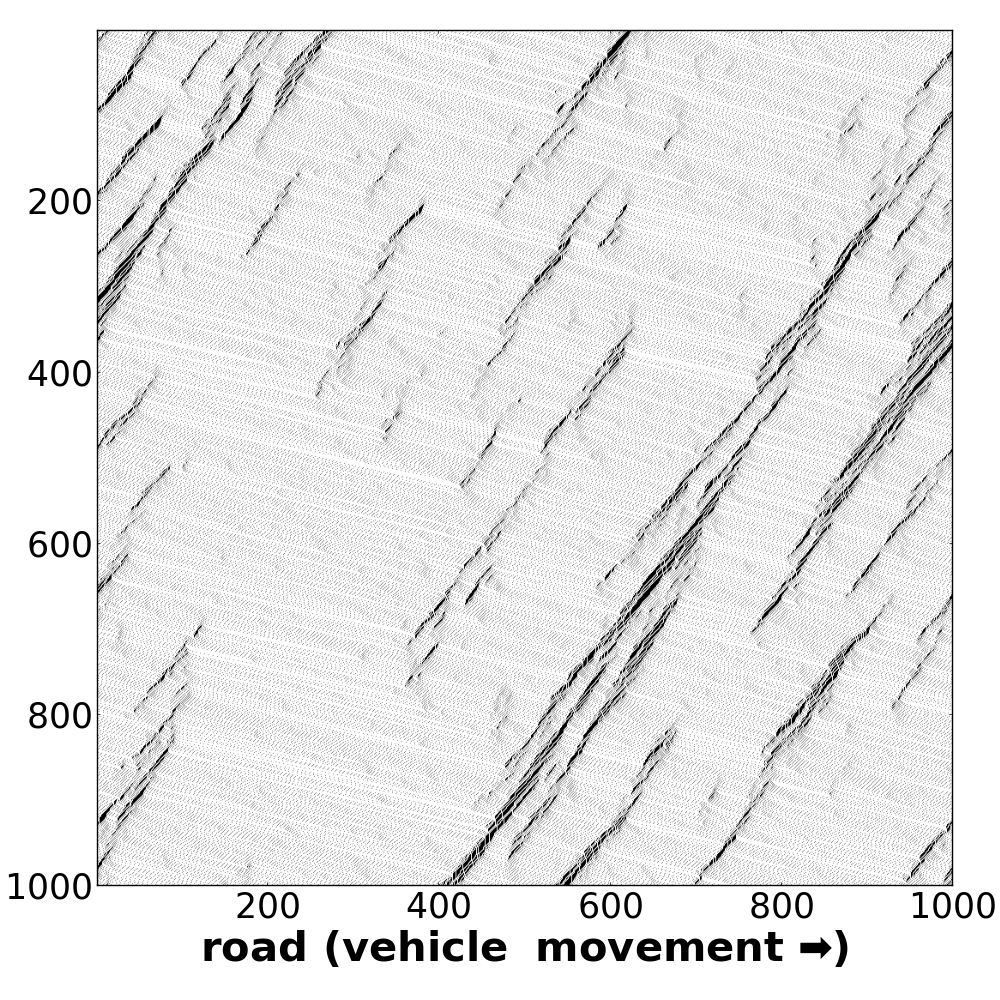}
                \caption{$p_{\text{brake}}=0.2$, $\rho=0.18$,\\$20 \% $ empowered agents}
                \label{20agents_0.2}
            \end{subfigure}
            &
            \begin{subfigure}{0.3\linewidth}
                \includegraphics[width=\linewidth]{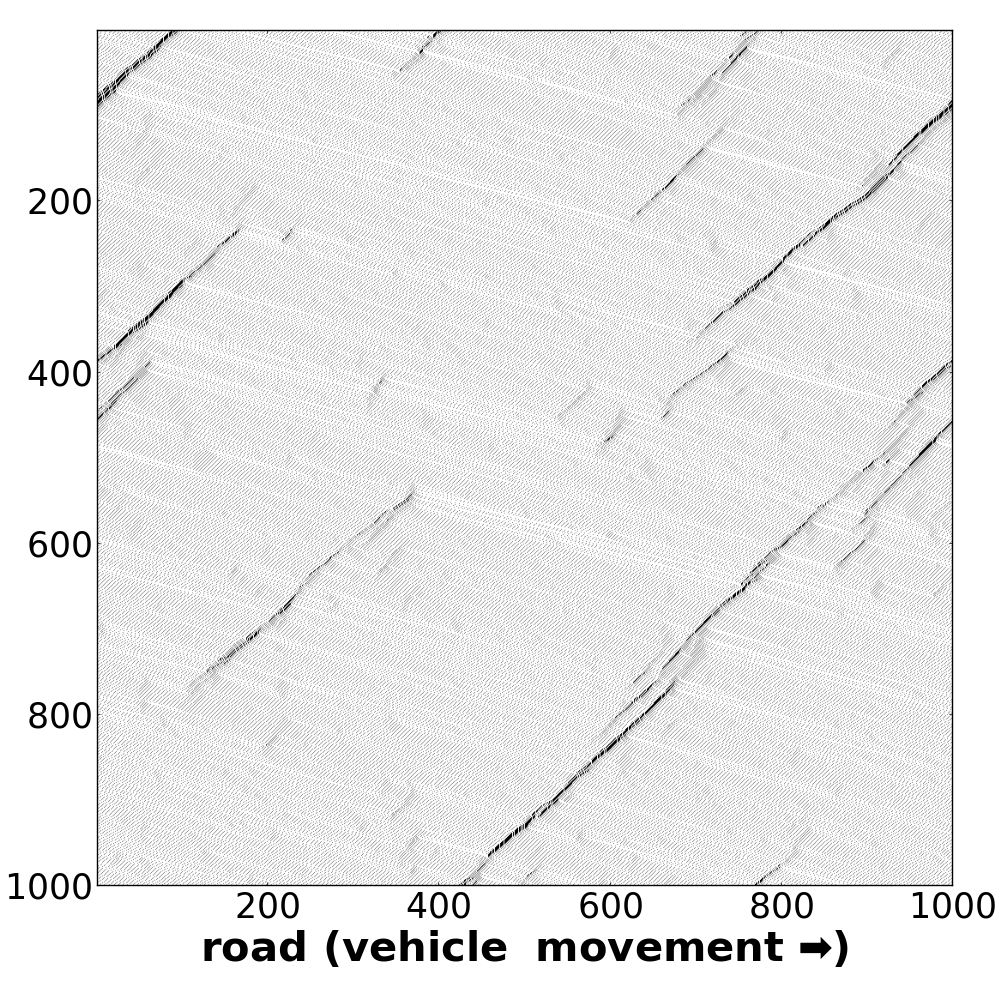}
                \caption{$p_{\text{brake}}=0.2$, $\rho=0.18$,\\$70 \% $ empowered agents}
                \label{70agents_0.2}
            \end{subfigure}
        \end{tabular}
        \\
        \begin{tabular}{ccc}
            \begin{subfigure}{0.3\linewidth}
                \includegraphics[width=1.05\linewidth]{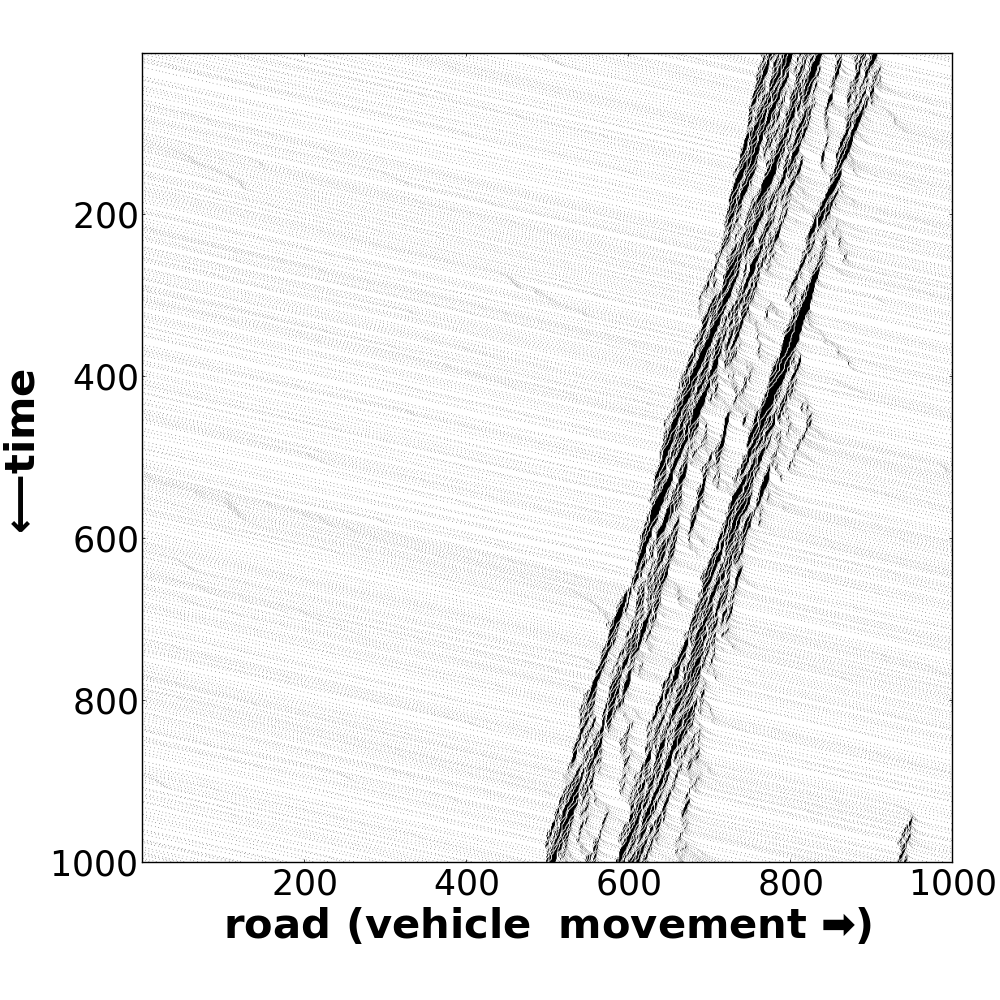}
                \caption{$p_{\text{brake}}=0.5$, $\rho=0.12$,\\NaSch baseline}
                \label{0agents_0.5}
            \end{subfigure}
            &\quad
            \begin{subfigure}{0.3\linewidth}
                \includegraphics[width=\linewidth]{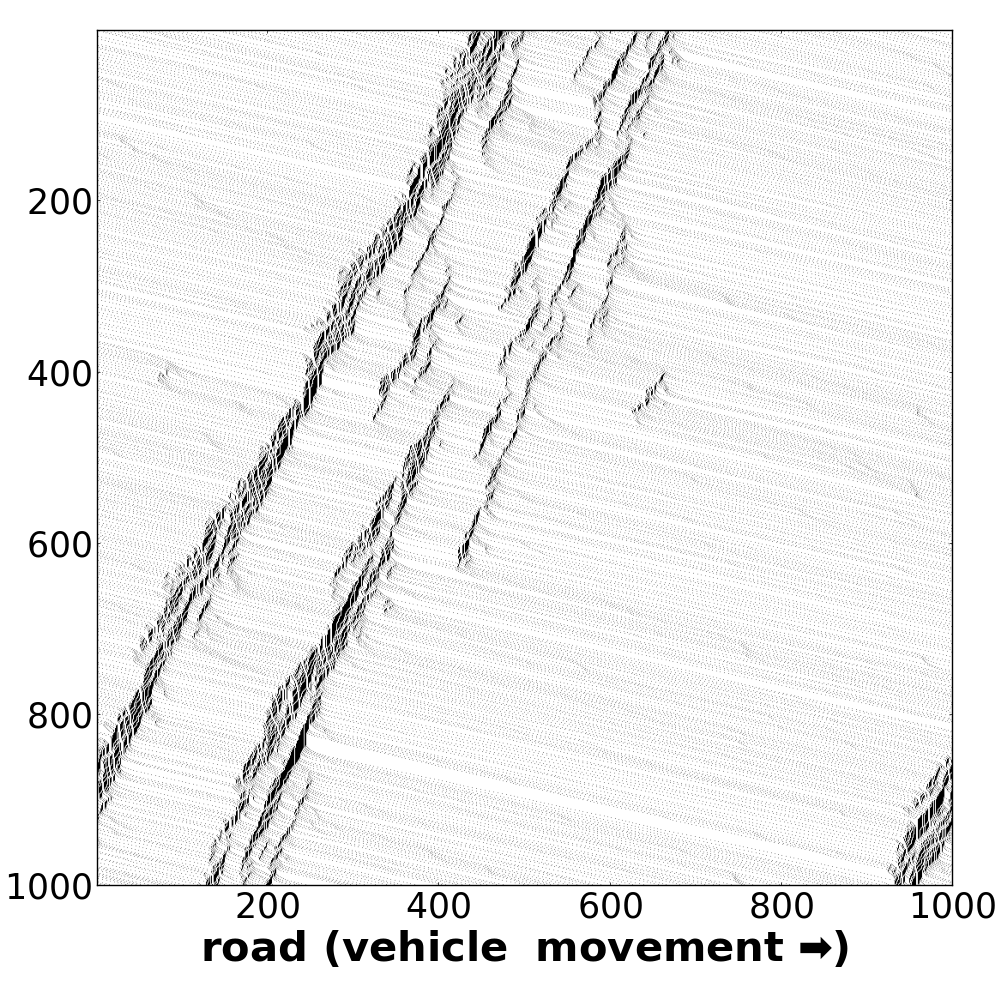}
                \caption{$p_{\text{brake}}=0.5$, $\rho=0.12$,\\$20 \%$ empowered agents}
                \label{20agents_0.5}
            \end{subfigure}
            &
            \begin{subfigure}{0.3\linewidth}
                \includegraphics[width=\linewidth]{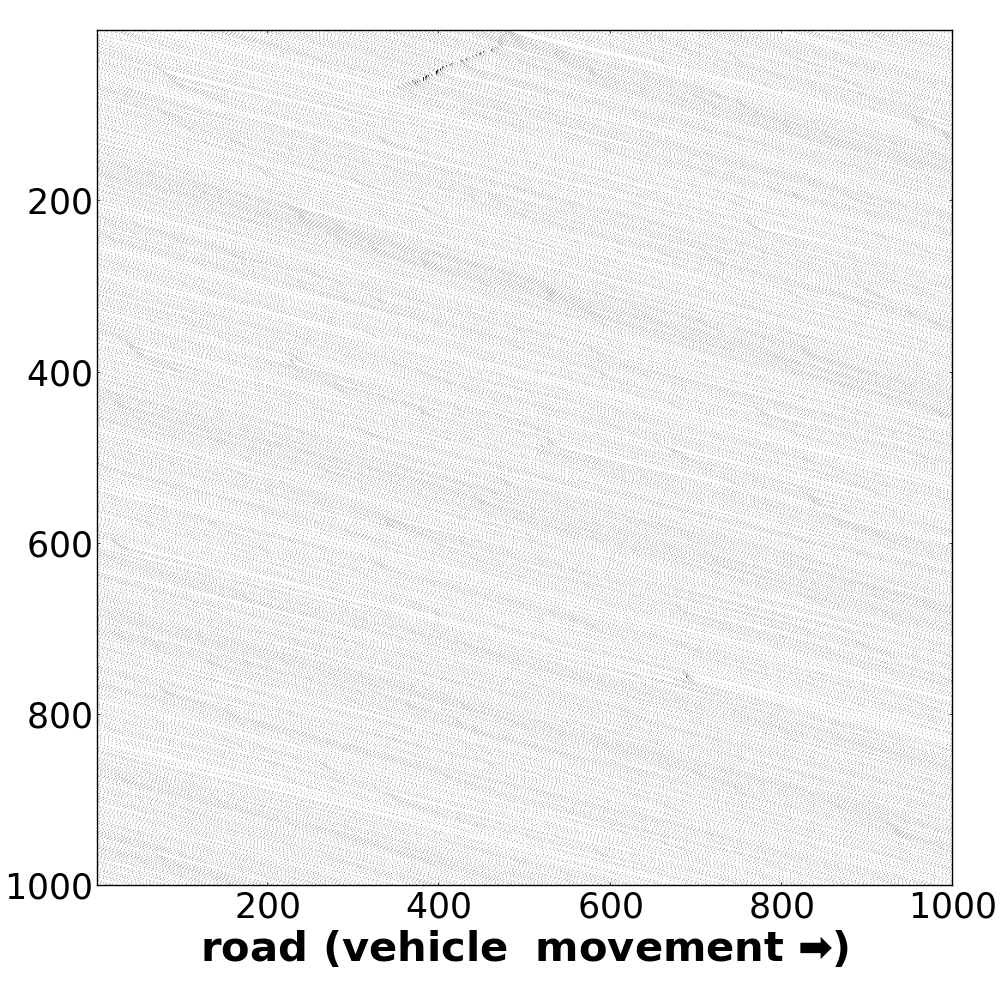}
                \caption{$p_{\text{brake}}=0.5$, $\rho=0.12$,\\$70 \% $ empowered agents}
                \label{70agents_0.5}
            \end{subfigure}
        \end{tabular}
    \end{tabular}
    \caption{Spatio-temporal view of traffic at densities corresponding to peak flow improvement achieved by our model. Each black dot represents a vehicle moving from left to right. The black streaks depict traffic jam waves propagating in reverse through the traffic. Increasing the number of empowered vehicles results in less dense jams and their quicker dissipation.}
    \label{time_space_plots}
\end{figure*}

The impact of empowerment is inherently local, constrained by the horizon. As depicted in Figure \ref{traffic flow with emp}, extending the empowerment horizon expands the range of densities where our model maintains effectiveness, illustrating its potential to mitigate traffic congestion across broader scenarios. The slight reduction of flow in low-traffic density (before {\it critical density $(\rho_c)$}) can be
intuitively explained by the fact that a competent driver never operates at the limits of their vehicle's performance, but leaves themselves options to react. As a consequence, they won't drive in a maximally aggressive fashion.

\begin{figure}[h!]
	\centering
	\includegraphics[width=0.7\columnwidth]{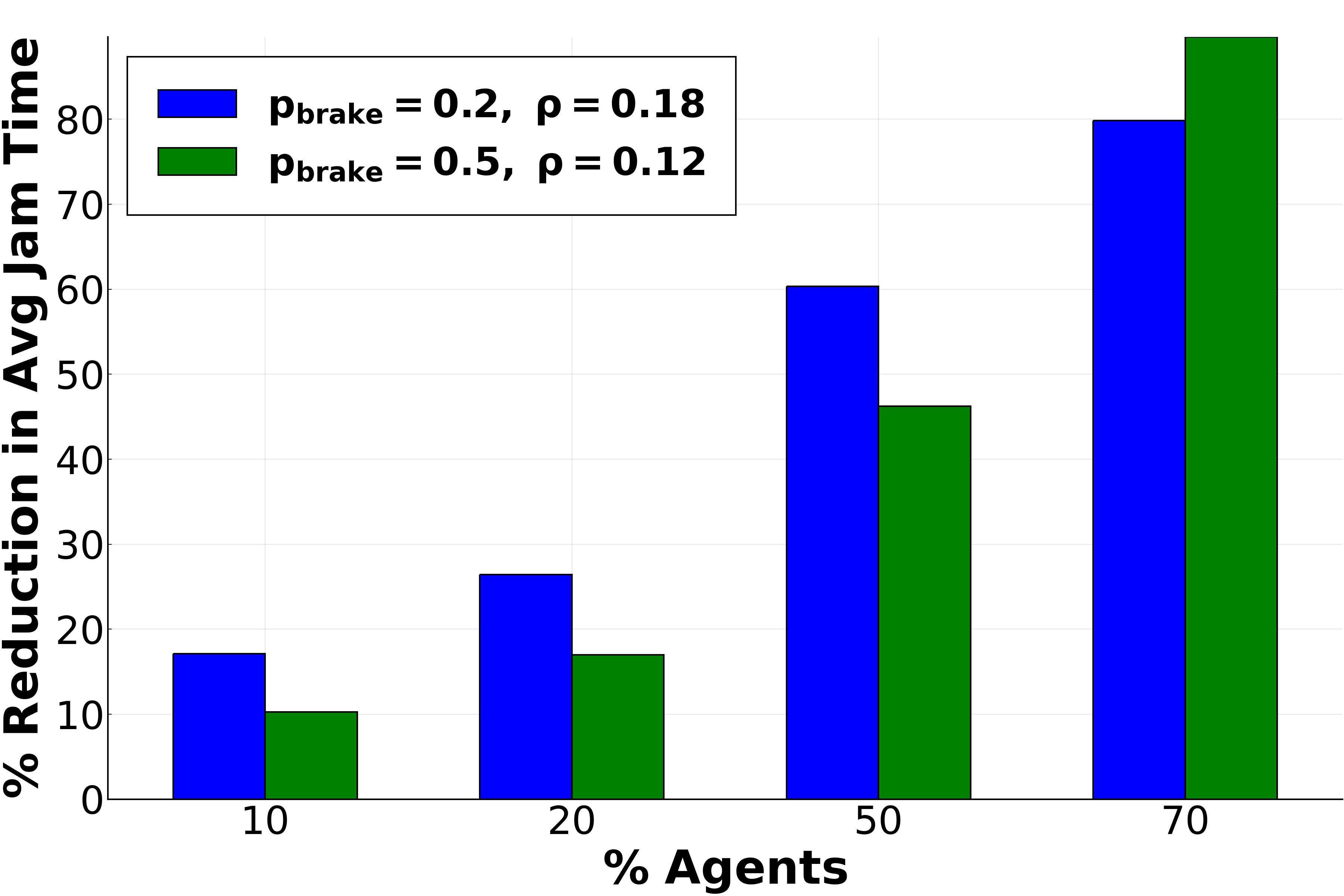} 
	\caption{ Percentage (\%) reduction in average traffic jam time for different ratios of empowered agents compared to a baseline model with varying $p_{\text{brake}}$ values at densities corresponding to peak flow improvement achieved by our model with a 3-step horizon.}
	\label{percentage reduction in jam time}
\end{figure}
For a deeper investigation into the factors driving improved flow in intermediate densities, we present the spatio-temporal diagram of vehicle movement in Figure~\ref{time_space_plots}. This figure offers a comprehensive view of vehicle dynamics over time, specifically focusing on densities where our model achieved peak flow improvement with a 3-step horizon, across different $p_{\text{brake}}$ values.
The horizontal axis depicts the road, which repeats periodically at the edges, hosting both normal cars and agents. The vertical axis indicates the  time (flowing from top to bottom). The cars are represented as black dots in space and time (no distinction is made between NaSch and empowered cars). The black streaks in the diagram now indicate traffic jam waves which, as NaSch remarkably captures, flow backwards through the traffic.

The comparison of the diagrams shows that  setups without agents (Figure~\ref{0agents_0.2},~\ref{0agents_0.5}) contain more concentrated and prolonged traffic jams indicating more number of cars trapped in jammed states for longer durations.
On the other hand, scenarios with empowered agents (Figure~\ref{20agents_0.2}, ~\ref{70agents_0.2}, ~\ref{20agents_0.5}, ~\ref{70agents_0.5}) result in shorter traffic jams involving fewer cars that are distributed more thinly across the road. This observation is corroborated by the significant decrease (up to $\approx$ 85\% ) in average jam time as depicted in Figure~\ref{percentage reduction in jam time} , corresponding to an increase in the percentage of empowered agents. In the computation of average traffic jam time, we followed the methodology outlined in \cite{nagel1994life}. The definition of jam time pertains to the number of time steps during which cars remain stationary. Specifically, cars with a velocity of $v=0$ are identified as stationary vehicles within our traffic model.

The effectiveness of empowered agents in mitigating congestion can be attributed to their adaptive braking behavior, which adjusts based on their current speed. Unlike conventional cars that brake with a fixed probability  $p_{brake}$ regardless of their speed (update {\it Rule 3}), empowered agents refrain from braking when traveling at low speeds, particularly when
stationary $(v = 0)$. The constant braking of regular cars, especially when stationary, contributes to the delay in jam dissolution. As long as there is sufficient leeway in the density, the tendency of the
empowered agents to maintain their own freedom of operation
appears to buffer off the fluctuations induced by the regular randomly
braking drivers.  As the density increases, the incentive of empowerment does
at a point no longer align with the goal of moving faster.
However, longer horizons can postpone the density where
the inflection happens. As seen in Figure~\ref{traffic flow with emp}(a, b), increasing
the empowerment horizon to 3 steps extends the range of
densities where empowerment improves flow

\section{Conclusion and Future Work}

We introduced and  examined a decentralized
strategy aimed at optimizing traffic flow. Our approach diverges substantially from traditional
decentralized approaches that often require the meticulous design
of reward functions tailored to the application, and instead
uses a generic model for intrinsic motivation, namely
\emph{empowerment}, to produce local behavior that improves global
traffic flow. Over a range of traffic densities, empowered cars can
substantially improve global traffic flow despite genuinely local
decision-making requiring no joint protocols and only
basic traffic rule coordination. 

The approach eliminates the need for explicit reward function
crafting and demonstrates another promising application of generic
intrinsic motivation approaches to address specific problems even in
multiagent scenarios. Notably, our solution
{\it does not} consist of adding hand-crafted rules to improve traffic flow. Rather, we show that the empowerment formalism
(which was already shown to create useful behaviours in a  large variety of other scenarios) also works almost out-of-the-box for the present scenario of decentralized traffic control.
The only model-specific assumption is essentially what part
of the state space to capture.

As a cellular automaton, the traditional NaSch model discretizes roads and velocities. Studying a continuous limit of the NaSch model \cite{krauss1996continuous} will require the generalization of empowerment towards the continuum and may reveal completely novel phenomena which we defer to future work.

NaSch has been a highly successful model for the study of typical traffic phenomena; crossings and many other extensions have been introduced  \cite{lakouari2018simulation, jiang2006effects,  pedersen2002entry}. 
For full deployment, it will be important to scale the method developed here not only towards the continuum, but towards crossings, various types of roads, and much more. 
Apart from the continuous limit, it would be of high interest to consider a hybrid model, where decentralized
control between the cars (either human or agents) is additionally globally coordinated by e.g., intelligent traffic lights and other elements. This requires a hierarchical approach interleaving centralized and decentralized control at different levels of the hierarchy. 

\section{Acknowledgments}

D.P. would like to thank Lukas Everding for numerous valuable discussions during an earlier pilot of the present. work. S.T. acknowledges support from NSF Grant No. 2246221 and Alliance Innovation Lab - Silicon Valley. H.P. was supported by the Division of Research and Innovation at San Jose State University (SJSU) under Award No. 23-SRF-03-041. The content is solely the responsibility of the authors and does not necessarily represent the official views of SJSU.

\bibliographystyle{IEEEtran}
\bibliography{IEEEexample.bib}

\end{document}